\documentclass[%
	 aip,
	 jap,%
	 amsmath,amssymb,
	reprint,%
	]{revtex4-1}
	
	\pdfoutput=1
	
	\usepackage{makecell} 
	\usepackage{booktabs} 
	\usepackage{mdframed}
	\usepackage{multirow} 
	\usepackage{graphicx}
	\usepackage{subfigure} 
	\usepackage{epstopdf}
	\usepackage{dcolumn}
	\usepackage{bm}
	\usepackage{subfigure}
	\usepackage[mathlines]{lineno}
	\def\sam{\textrm{sam}}
	\def\ins{\textrm{ins}}

	\def\resid{\textrm{resid}}
	\def\heater{\textrm{heater}}
	\def\th{\textrm{th}}
	\def\total{\textrm{total}}
	\def\source{\textrm{source}}

\begin{document}

        
	\title[Modulation calorimetry in DACs II]{Modulation calorimetry in diamond anvil cells II: Joule-heating design and prototypes}
	\author{Zachary M. Geballe}
	\affiliation{Geophysical Laboratory, Carnegie Institution for Science, Washington, DC 20015}
	\author{Viktor V. Struzhkin}
	\affiliation{Geophysical Laboratory, Carnegie Institution for Science, Washington, DC 20015}
	\author{Andrew Townley}
	\affiliation{Department of Electrical Engineering and Computer Sciences, University of California, Berkeley, CA 94720}
	\author{Raymond Jeanloz}
	\affiliation{Department of Earth and Planetary Science, University of California, Berkeley, CA 94720}
	\date{\today}	

	\begin{abstract}
	Part I shows that quantitative measurements of heat capacity are theoretically possible inside diamond anvil cells via high-frequency Joule heating (100 kHz to 10 MHz), opening up the possibility of new methods to detect and characterize transformations at high-pressure such as the glass transitions, melting, magnetic orderings, or the onset of superconductivity. Here we test the possibility outlined in Part I, using prototypes and detailed numerical models. First, a coupled electrical-thermal numerical model shows that specific heat of metals inside diamond cells can be measured directly using $\sim 1$ MHz frequency, with $< 10\%$ accuracy. Second, we test physical models of high-pressure experiments, i.e. diamond-cell mock-ups. Metal foils of 2 to 6 $\mu$m-thickness are clamped between glass insulation inside diamond anvil cells. Fitting data from 10 Hz to $\sim 30$ kHz, we infer the specific heat capacities of Fe, Pt and Ni with $\pm 20$ to $30\%$ accuracy. The electrical test equipment generates -80 dBc spurious harmonics which overwhelm the thermally-induced harmonics at higher frequencies, disallowing the high precision expected from numerical models. An alternative Joule-heating calorimetry experiment, on the other hand, does allow absolute measurements with $< 10\%$ accuracy, despite the -80 dBc spurious harmonics: the measurement of thermal effusivity, $\sqrt{\rho c k}$ ($\rho$, $c$ and $k$ being density, specific heat and thermal conductivity), of the insulation surrounding a thin-film heater. Using a $\sim 50$ nm-thick Pt heater surrounded by glass and 10 Hz to 300 kHz frequency, we measure thermal effusivity with $\pm 6\%$ accuracy inside the sample chamber of a diamond anvil cell.
	\end{abstract}
		
	\newcommand{\ud}{\mathrm{d}}
          \maketitle
\section{Introduction}
High-frequency calorimetry of metal samples in diamond anvil cells has the potential to reveal Debye temperatures, deviations from Debye models, heat capacity anomalies at magnetic, superconducting, or amorphization transitions, and the latent heats of melting and other first-order transitions. Such measurements would complement existing structure-sensitive high-pressure techniques (e.g. x-ray diffraction) and enable comparison of diamond-cell data with shock-wave data that are intrinsically adiabatic, but which operate under different conditions (e.g. short time scales, high strain rates, irreversibility of pressure-temperature paths).

The primary challenge in such an experiment is to heat a small sample in a nearly-adiabatic manner despite the fact that it is bordered by a solid or liquid of high thermal conductivity ($k \sim 1$ to 30 Wm$^{-1}$K$^{-1}$) and is within $\sim 10$ $\mu$m of diamond anvils ($k \sim$ 2000 Wm$^{-1}$K$^{-1}$), yielding a thermal diffusive timescale of $\sim 10$ to 100 $\mu$s.

In fact, this challenge is also encountered in measurements of materials grown on thermally conductive substrates such as Si or Al$_2$O$_3$, meaning the results and analysis presented here may facilitate measurements in applications outside high-pressure research. In particular, if an as-grown material is $> 1$ $\mu$m thick, it is amenable to the same high-frequency calorimetric measurements studied here, without the need for high-frequency modulated lasers and photodiodes (as in Refs. \onlinecite{Cahill2004,Wei2013}). 

Within high-pressure experimental science, a few pioneering methods have been employed to study heat capacity in-situ. The highest pressure experiments\cite{Demuer2000,Fernandez-Panella2011,Baloga1977,Bouquet2000,Sidorov2011,Sidorov2013} have used laser heating or resistive heating at frequencies up to hundreds of Hz or even 10 kHz in one case, samples ranging from nanoliters to microliters in volume, and maximum pressures and temperatures up to 13 GPa and 20 K in one case\cite{Fernandez-Panella2011} and 0.3 GPa and 300 K in another.\cite{Baloga1977} 

To make quantitative measurements at higher pressures and temperatures (10 GPa and 300 K to 100 GPa and 3000 K), Part I of this two-part publication shows that even higher frequencies (kHz to MHz) are required, and that Joule-heating enables absolute measurement of specific heat. Here we extend Joule-heating modulation calorimetry to higher frequencies and smaller sample sizes than previously achieved, using two methods to study measurement accuracy: a detailed electrical-thermal model of the design introduced in Part I, and laboratory measurements of metal heaters ranging in volume from $\sim 4$ to 40 picoliters that are pressed between glass insulation inside diamond anvil cells. 

In both the numerical model and laboratory measurements, power is deposited via Joule heating of metal foils and temperature oscillations are measured via the third harmonic technique \cite{Kraftmakher2004} using a bridge circuit adapted from one used at ambient pressure for specific heat spectroscopy.\cite{Birge1997}


\begin{figure*}
\includegraphics[width=6.5in]{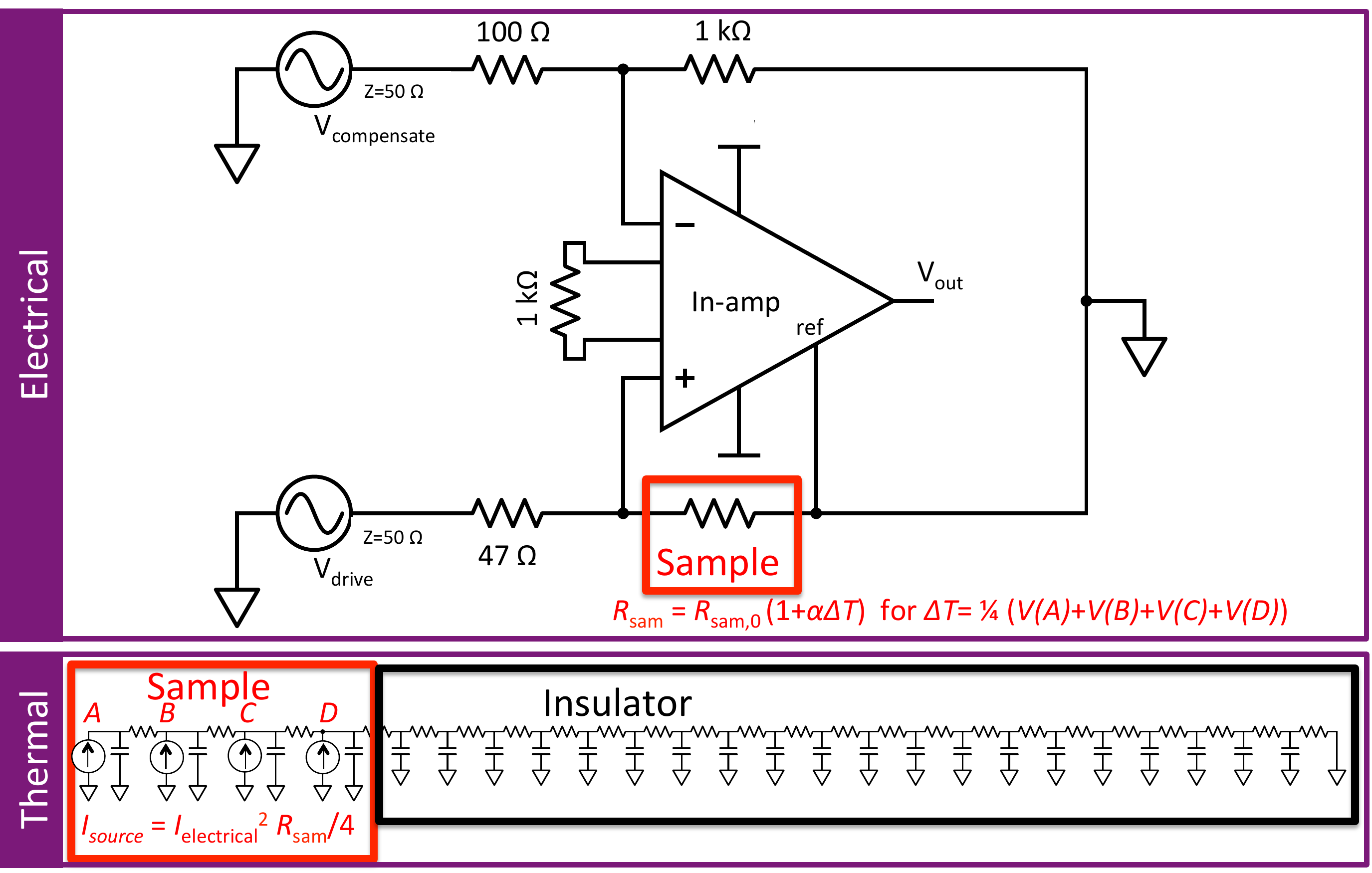} 
\caption{Schematic of the coupled electrical-thermal model. (Top) Electrical design of bridge circuit, with red box highlighting sample resistor. (Bottom) One-dimensional thermal model of half the sample thickness and one side of insulation, where electrical symbols are used to model heat flow (as electrical current) and temperature (as voltage).}
\label{fig:electrical-thermal}
\end{figure*}

\section{Numerical Method}
First, we test whether a coupled electrical-thermal model predicts the same results as the thermal model of Part I, in which temperature was simulated and used to estimate electrical resistance, but with no feedback from resistance to heating power. \footnote{See www.analog.com/en/products/amplifiers/instrumentation-amplifiers/ad8421.html.} These coupled models test a few key assumptions implicit in Part I: (1) that the amplitudes of currents and voltages needed to induce measurable third harmonic voltage oscillations are in the typical range available from commercial test equipment, (2) that the resistance oscillations in our design are small enough to use the approximation ``$\frac{1}{1+x} \approx 1-x$'' to infer heat capacity in Appendix C, and (3) that harmonic distortions from the instrumentation amplifier do not bias the third-harmonic temperature measurement.

A schematic of our coupled electrical-thermal model is shown in Fig. \ref{fig:electrical-thermal}. Two outputs of a waveform generator drive voltage oscillation through the two arms of a bridge circuit, one of which contains the metal sample. The driving voltage in the sample arm is,
\[ V_d \sin(\omega t) 
\]
The generator's other output sends a compensating voltage, $V_c \sin(\omega t)$, through two resistors, with an amplitude that is tuned so that the voltage across the 1 k$\Omega$ reference resistor equals the main component of voltage across the sample. We implement this model in LTSpice.\footnote{See www.linear.com/LTspice}

\begin{figure*}
\includegraphics[width=6in]{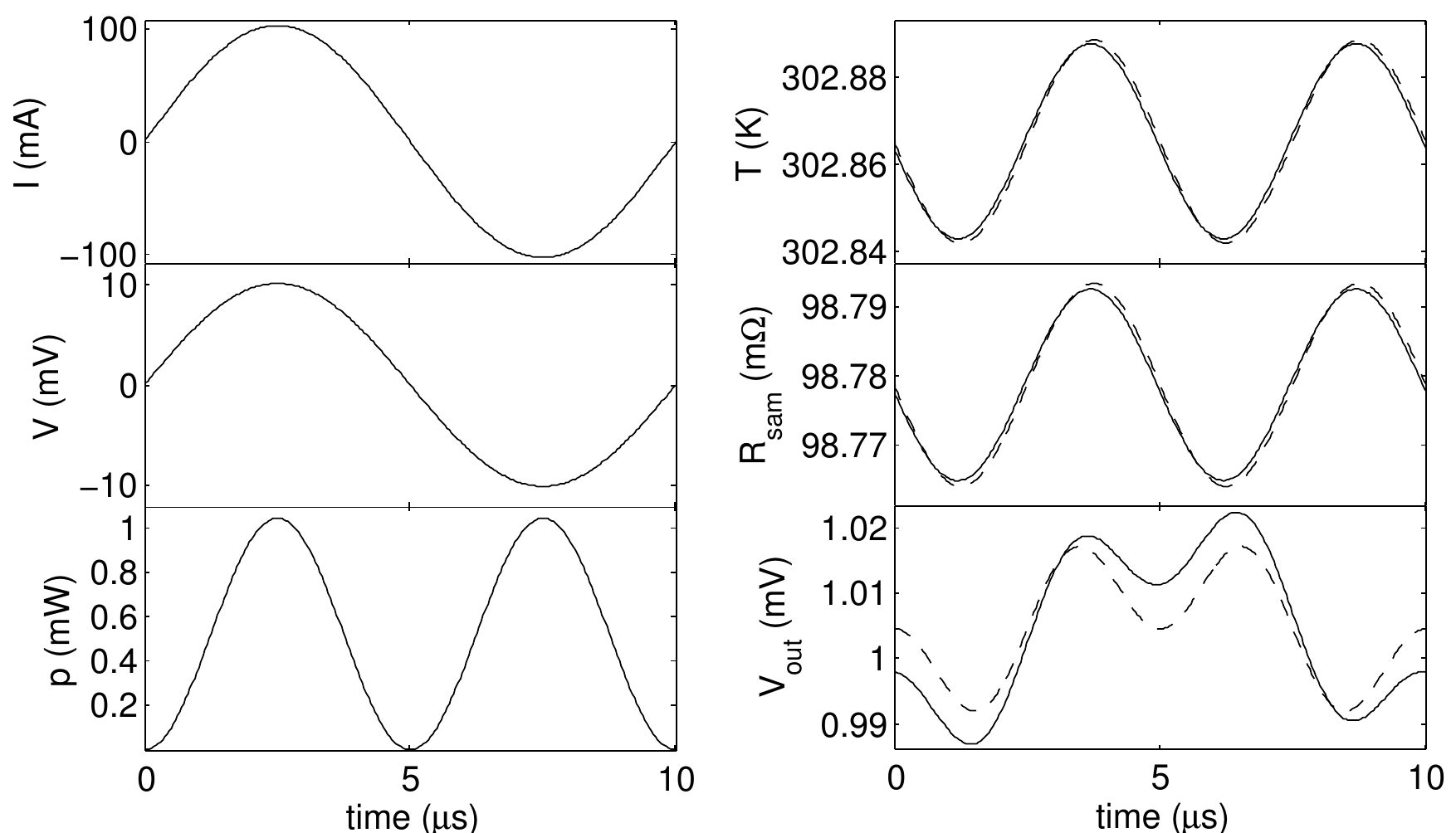} 
\caption{Result of numerical model at 100 kHz frequency. From top-left, a 100 kHz, 100 mA alternating current drives 10 mV voltage oscillations in the sample, creating a 200 kHz power oscillation from 0 to 1 mW, which causes the average sample temperature to oscillate from 0.69 to 0.73 K above room temperature, causing resistance to oscillate from 97.76 to 97.79 $m\Omega$. This dynamic resistance feeds back into voltage oscillations across the sample, which can be measured with the help of a bridge circuit (Fig. \ref{fig:electrical-thermal}) that nulls out the dominant signal due to the average value of resistance. The residual voltage is shown in the bottom-right panel. It is the output of an in-amp, which amplifies the difference in voltage across the bridge, with a gain of 10.9. It also distorts the signal and adds a DC offset, but the in-amp does not generate significant spurious voltages at the third harmonic frequency. Dashed curves show what the temperature, resistance and residual voltages would be in the ideal case of no addenda contribution and ideal electronics.}
\label{fig:3w_LTSpice}
\end{figure*}

At ideal tuning, the bridge is ``balanced'' and most of the voltage at the inverting input of the in-amp is ``nulled out'' by the compensating voltage. An example illustrates tuning of the bridge; Fig. \ref{fig:3w_LTSpice} shows that at 100 kHz frequency and $\pm 4.8$ V driving voltage, the voltage difference between midpoints of the two arms of the bridge can be minimized by balancing the bridge, resulting in a mustache-shaped waveform at the output of the in-amp (final frame of Fig. \ref{fig:3w_LTSpice}). Appendix A describes the calculation of compensating voltage needed to balance the bridge. Alternatively, balance can be achieved by trial and error.

To model the temperature oscillation, the LTSpice electrical software is used once again. This time, electrical components are used to make the elements of a finite element model that matches the one used in Part I of this two-part publication, with one exception: LTSpice's native time-stepping routine is used instead of the Crank-Nicholson scheme used previously. Material properties and dimensions of the sample assumed here are identical to those modeled in Part I, matching the properties of iron at ambient conditions. The insulation material is assumed to have the properties of silica glass at ambient conditions, which is slightly less thermally conductive than the KBr insulation modeled in Part I. We choose to model a different insulating material here in order to decrease the addenda contribution to total heat capacity and to enable comparison with laboratory tests using a metal film deposited on glass (see below).

 Fig. \ref{fig:electrical-thermal} outlines the thermal model used in LTSpice. The sample is divided into four elements, each of which is heated by a current source equal to one-forth the electrical power deposited in the sample. This flow of heat increases the temperature of the sample elements, which are linked via thermal resistances, and connected to a sequence of twenty insulator elements that are also linked via thermal resistances. \footnote{Each element represents a layer of material with thickness $dz = \frac{5}{8}$ $\mu$m, providing a coarse mesh to model half the thickness of a 5 $\mu$m-thick piece of metal. The coarseness of the mesh results in $\sim 10\%$ numerical error (evidenced by the discrepancy compared to results using the fine mesh of Matlab simulations in Part I). To compensate, we average two model runs using different assumptions of conductance between metal and insulator, as described in Appendix B.} Correspondences between thermal parameters and electrical parameters used in the computer model are listed in Table \ref{table:electrical-thermal}, and details are given in Appendix B.

The thermal and electrical models of the sample are coupled in the following way. Electrical power across the electrical model of the sample causes heat flow in the thermal model of the sample: $dQ/dt = I_{\sam}V_{\sam}/4$, where $I_{\sam}$ and $V_{\sam}$ are values of current across the sample, $Q$ is the heat added to each sample element, and the 4 accounts for the four sample elements. The average temperature of the four thermal sample elements, $T$, modulates the resistance of the electrical model of the sample: $R_{\sam} = R_0(1+\alpha T)$ where $R_0 = r\cdot l/w\cdot d$ is the room temperature sample resistance and $\alpha = d\textrm{log}r/dT$ is the temperature coefficient of resistance.

\begin{table}
\begin{center}
\begin{tabular}{ p{4 cm} c c }  
& Sample (Fe) & Insulator (silica glass) \\
\hline 
$d$: Layer thickness ($\mu$m) & 5 & 12.5 \\
$w$: Width ($\mu$m) & 20 & 20 \\
$l$: Length ($\mu$m) & 100 & 100 \\
$\rho$: Density (g cm$^{-3}$) & 7.9 & 2.2 \\
$c$: Specific heat (J g$^{-1}$ K$^{-1}$) & 0.45 & 0.83 \\
$k$: Thermal conductivity (W m$^{-1}$ K$^{-1}$) & 80 & 1.2 \\
$r$: Resistivity ($\Omega$ m) & $9.7 \times 10^{-8}$ & 0 \\
$\alpha$: Temperature coefficient of resistance (K$^{-1}$) & 0.0064 & 0 \\
\hline 
\end{tabular}

\caption{Properties of the sample and insulator used in our numerical models.}
\label{table:mat_props_v3}
\end{center}
\end{table}

\begin{table}
\begin{center}
\begin{tabular}{ p{4 cm} @{\hspace{1 cm}} p{3.2 cm}} 
Thermal & Electrical \\
\hline
power, $p$(W) & current, $I$(A) \\
heat, $Q$(J) & charge, $Q$(C) \\
temperature, $T$(K) & voltage, $V$(V) \\
heat capacity, $c_\sam \rho_\sam A \Delta z$ (J/K) & capacitance, $C$(C$^2$/J) \\
thermal resistance, $\Delta z/kA$ (K/W) & resistance, $R$($\Omega$) \\ 
\hline
\end{tabular}

\caption{Thermal parameters and the electrical parameters used to model them, including variable and units. Note that $A$ is the sample surface area, and $\Delta z$ is the thickness of each element in our finite element model.}
\label{table:electrical-thermal}
\end{center}
\end{table}

\begin{figure}
\begin{center}
\includegraphics[width=3.3in]{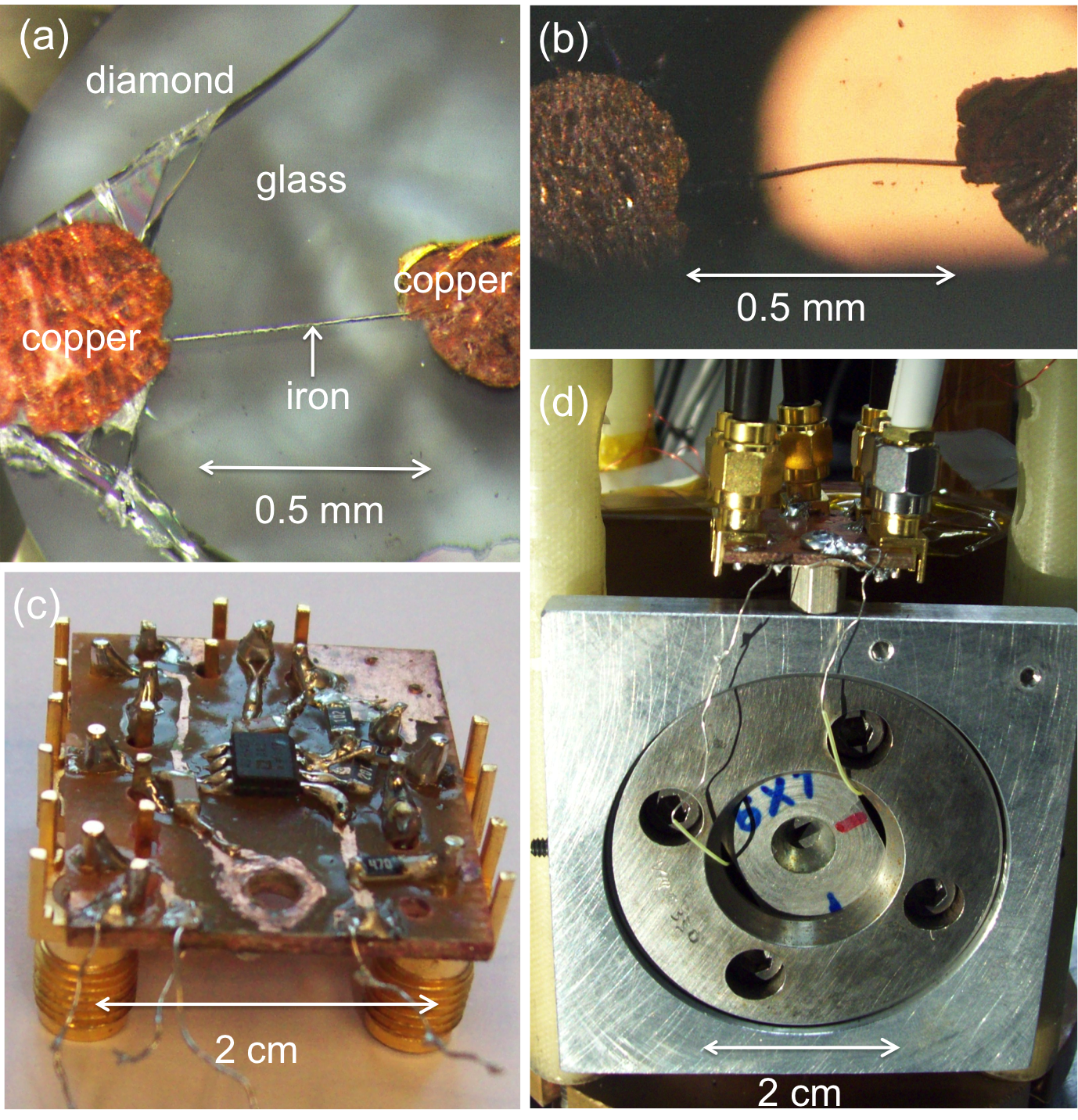} 
\caption{(a) A strip of iron resting on a 120 $\mu$m-thick piece of a glass on top of a diamond anvil, with copper leads resting on top of the iron. (b) The same materials as in (a), but after a second glass-covered diamond anvil has been pressed against the iron and copper. (c) The circuit board used to connect electrical test equipment (wavefunction generator, DC power supply and oscilloscope) to the iron sample. (d) Diamond cell connected to electrical board.}
\label{fig:GL_setup}
\end{center}
\end{figure}

\section{Laboratory Method}

To test whether our electrical and thermal design can be implemented in practice, we first study iron, platinum and nickel foils pressed inside diamond-cells at near-ambient pressure.\footnote{The large thickness of glass that separates the diamonds implies almost no pressure: assuming the yield stress of glass is $\sigma_Y \sim 25$ MPa (the bending strength of Schott's borofloat glass), then the maximum pressure is approximately 300 $\mu$m $\times \sigma_Y/ 240$ $\mu$ m $\sim 30$ MPa.} We make diamond-glass-metal-glass-diamond sandwiches using two pieces of microscope slide coverslips (120 $\mu$m-thick silicate glass) and 2.4, 5.7 and 6 $\mu$m-thick foil of platinum, iron and nickel, respectively. The iron and platinum samples are cut with razor blades while the nickel sample is laser-cut. By testing these circuits with electronics that mimic the schematic used in our numerical model (Fig. \ref{fig:electrical-thermal}), we determine whether our diamond-cell calorimetry design is feasible. Potential pitfalls include electrical noise, spurious harmonic distortion, contact resistance, or electromigration of heater material. We will show that spurious harmonic distortions limit our accuracy, but note that contact resistance did overwhelm the sought-after third harmonic during testing not presented here in the case of silver epoxy contacts cured at room temperature.

Second, we test a thin-film of platinum (50 nm thick) sputtered onto the central $16 \pm 1$ $\mu$m-wide region of 10 $\mu$m-thick glass disc, using photolithography. It is pressed against a second disc of glass (20 $\mu$m-thick) inside the sample chamber of a diamond-cell at near-ambient pressure. Heater length and width are measured with an optical microscope, while thickness is measured with a Zygo optical surface profiler.

The electrical test equipment is the same for both thin-film and foils, and follows the same design used in our numerical models (Fig. \ref{fig:electrical-thermal}). A 14-bit, 500 megasamples per second waveform generator (BK Precision 4065) delivers both drive and compensation sine waves via its two outputs. Resistors and the amplifier for the bridge circuit are soldered onto the homemade circuit board shown in Fig. \ref{fig:GL_setup}b. The amplifier is powered by $\pm 10$ V DC power, with 1 $\mu$F bypass capacitors to ground that filter out high-frequency noise. The output from the amplifier is read by an 8-bit 500 megasample per second oscilloscope with minimum sensitivity of 2 mV per division (Lecroy LT342), or by a 10 MHz lock-in amplifier (Zurich Instruments HF2LI). A third alternative is to avoid the in-amp and to compensate the sample's first harmonic voltage through the differential input of the lock-in amplifier instead. All three voltage-measuring schemes are tested, and the differences are seen to be negligible (section \ref{section:errors}).\footnote{The in-amp plus lock-in were used to collect data in Figs. \ref{fig:Fe_data} and \ref{fig:Pt_data}, the in-amp plus oscilloscope was used to collect data in Figs. \ref{fig:Ni_data}, \ref{fig:glass_data}, \ref{fig:glass_errors}, and both were used in Figs. \ref{fig:Fe_errors},\ref{fig:Pt_errors}.}

\begin{figure}
\begin{center}
\includegraphics[width=3.5in]{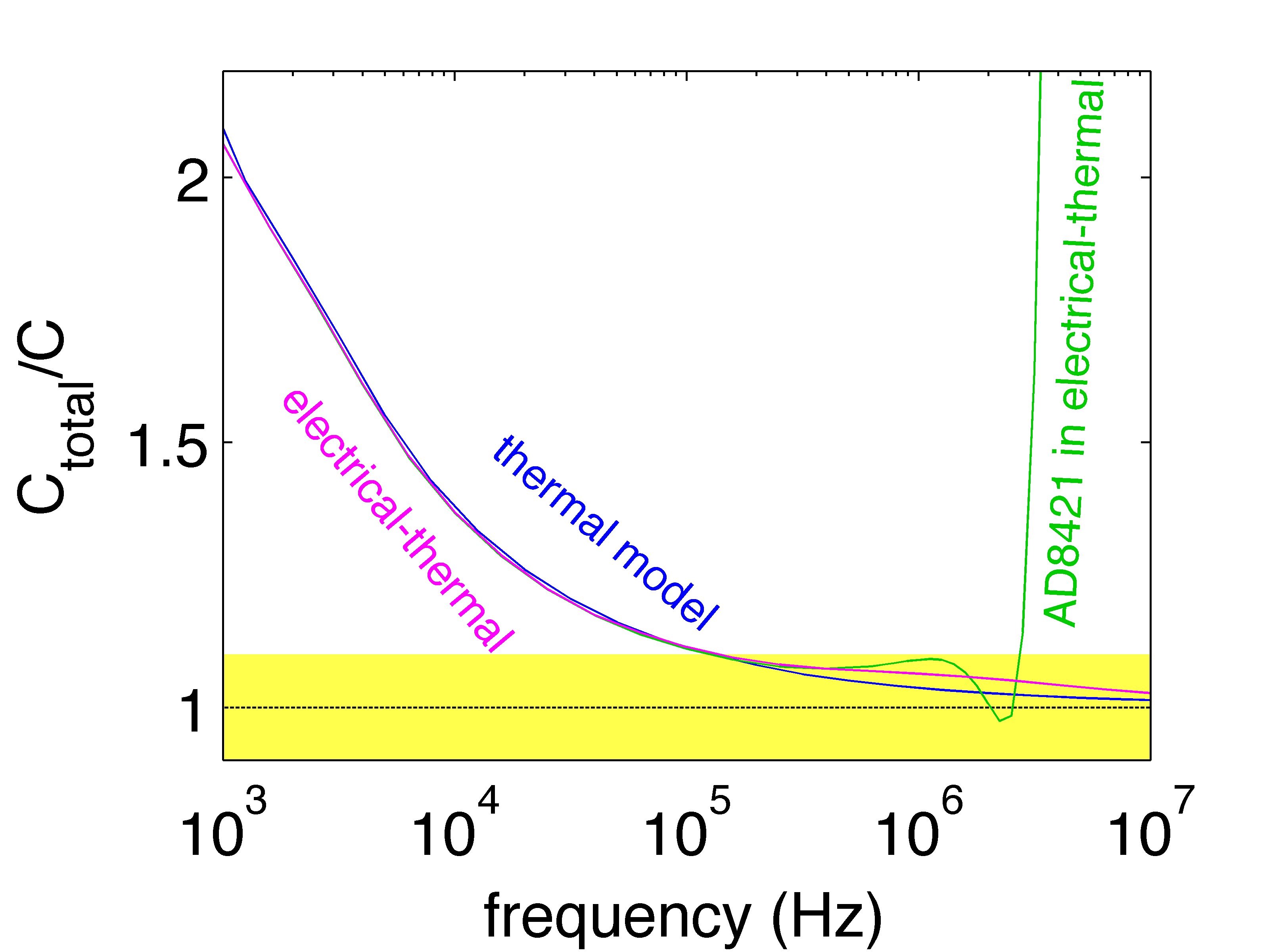} 
\caption{Total heat capacity that would be measured in numerical models, normalized by heat capacity of the sample alone, as a function of frequency. Colors indicate the one-dimensional finite element model that was used: thermal model described in Part I (blue), or the coupled electrical-thermal model described here with ideal models (pink) or manufacturer supplied models (green) of the in-amp. Yellow highlights the region of $< 10\%$ error. Note that frenquencies plotted in this paper refer to the fundamental mode of current and voltage oscillation, and are therefore two-times smaller than the heating frequencies referred to as ``frequency'' in Part I.}
\label{fig:CbyC_LTSpice}
\end{center}
\end{figure}

\section{Numerical Results}

Fig. 2 shows several consequences of a 100 kHz driving voltage for the iron heater modeled numerically. The quantity we measure in laboratory experiments is shown in the final figure: the amplitude of third harmonic voltage oscillation, $V_{3\omega}$. 

Appendix C shows that by measuring $V_{3\omega}$, along with the values of time-averaged resistance, $R_\sam$, and in-amp gain, $G$, we can experimentally determine the amplitude of temperature oscillations in the heater (also see Ref. \onlinecite{Birge1997}): 

\begin{equation}
T_{2\omega} = \frac{2 \left( R_b + R_\sam  \right)^2 V_{3\omega}}{\alpha V_d R_\sam R_b G} 
\label{eqn:T2w_meas}
\end{equation}

where $R_b = 97$ $\Omega$ is the buffer resistance (i.e. total resistance between voltage-generation and sample), $\alpha$ is the assumed or measured temperature coefficient of resistance, and $V_d$ is the driving voltage inside the waveform generator (i.e. the nominal voltage in the ``high Z'' mode of the BK 4065 waveform generator).

Since we also know the power deposited via Ohm's law, we can determine the heat capacity of the heater plus addenda (i.e. whatever nearby insulating material is dynamically heated):

\begin{eqnarray}
C_\total &=& p_{2\omega}/2\omega T_{2\omega} \nonumber \\
&=& \frac{1}{2} ( \frac{V_d}{R_\sam+R_b} )^2 R_\sam \frac{1}{2\omega T_{2\omega}} \nonumber \\
&=& \frac{ \alpha V_d^3 R_\sam^2 R_b G}{8 \omega (R_\sam + R_b )^4 V_{3\omega} }  \label{eqn:C_meas} 
\end{eqnarray}

We now apply these formulas to the numerically modeled third harmonic amplitudes. The temperature oscillation inferred from the measurement of third harmonic amplitude, $V_{3\omega} = 7.8$ $\mu$V, would be $T_{2\omega} = 22$ mK, which is 11\% smaller than what the temperature oscillation would be in a truly adiabatic experiment. The inferred heat capacity would be $C_\total = 39 \textrm{ nJ/K}$, which is 11\% larger than the 36 nJ/K heat capacity of the 79 ng mass (10 picoliter volume) iron sample modeled here. The discrepancy arises from heat loss to the addenda (i.e. thermal conduction into the insulation that increases the spatial extend, hence total heat capacity of the dynamically heated region).

The frequency dependence of this inferred heat capacity is shown in Fig. \ref{fig:CbyC_LTSpice}. At frequencies  $\leq 1$ MHz, the results of this coupled electrical-thermal model match those of the thermal model presented in Part I, confirming that the voltage oscillations resulting from calculated thermal oscillations approximate those in the more-realistic case of coupled voltage and thermal oscillations. The match also shows that at low-enough frequencies, a realistic in-amp provides a high-fidelity voltage output that can be used to infer heat capacity of a metal sample. At frequencies beyond 1 MHz, however, the in-amp distorts the third harmonic measurement, suggesting that at least with this differencing amplifier, high-fidelity third harmonic voltages cannot be extracted at the highest frequencies studied in Part I of this two-part publication.

\begin{figure}
\begin{center}
\includegraphics[width=3.5in]{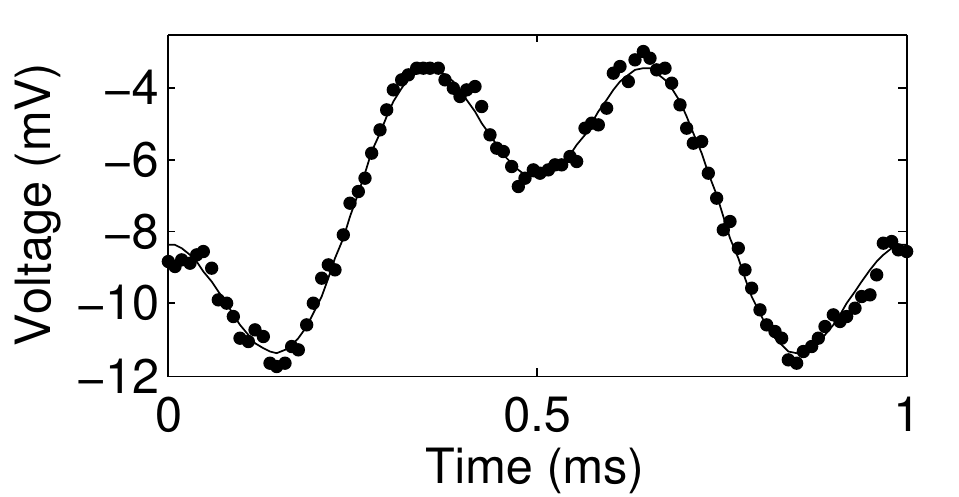} 
\caption{Measured voltage timeseries (dots) and curve fitted using first and third Fourier components. Five hundred 1 ms cycles were averaged to produce the timeseries.}
\label{fig:data_ex}
\end{center}
\end{figure}
\begin{figure}
\begin{center}
\includegraphics[width=3in]{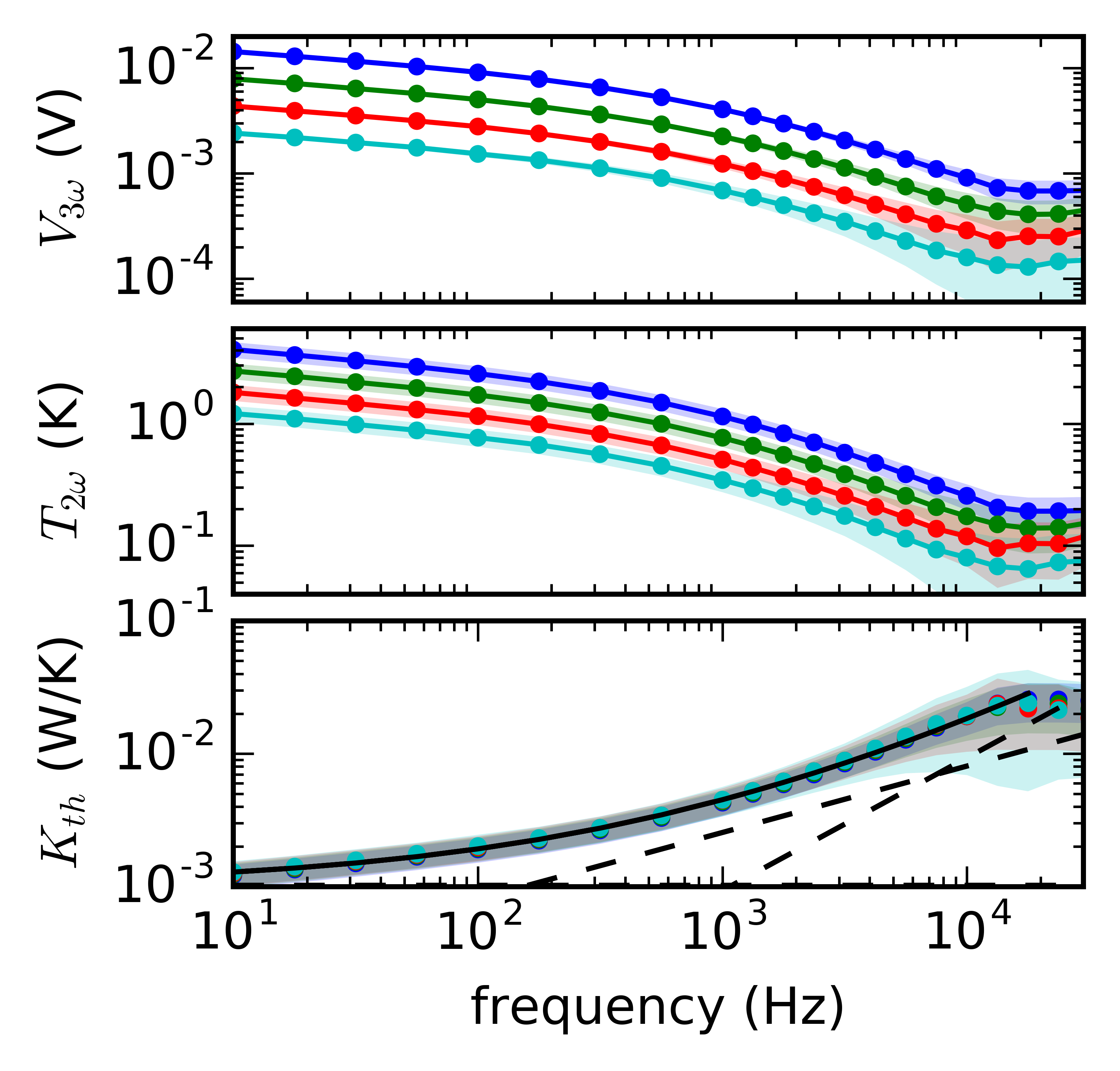} 
\includegraphics[width=3in]{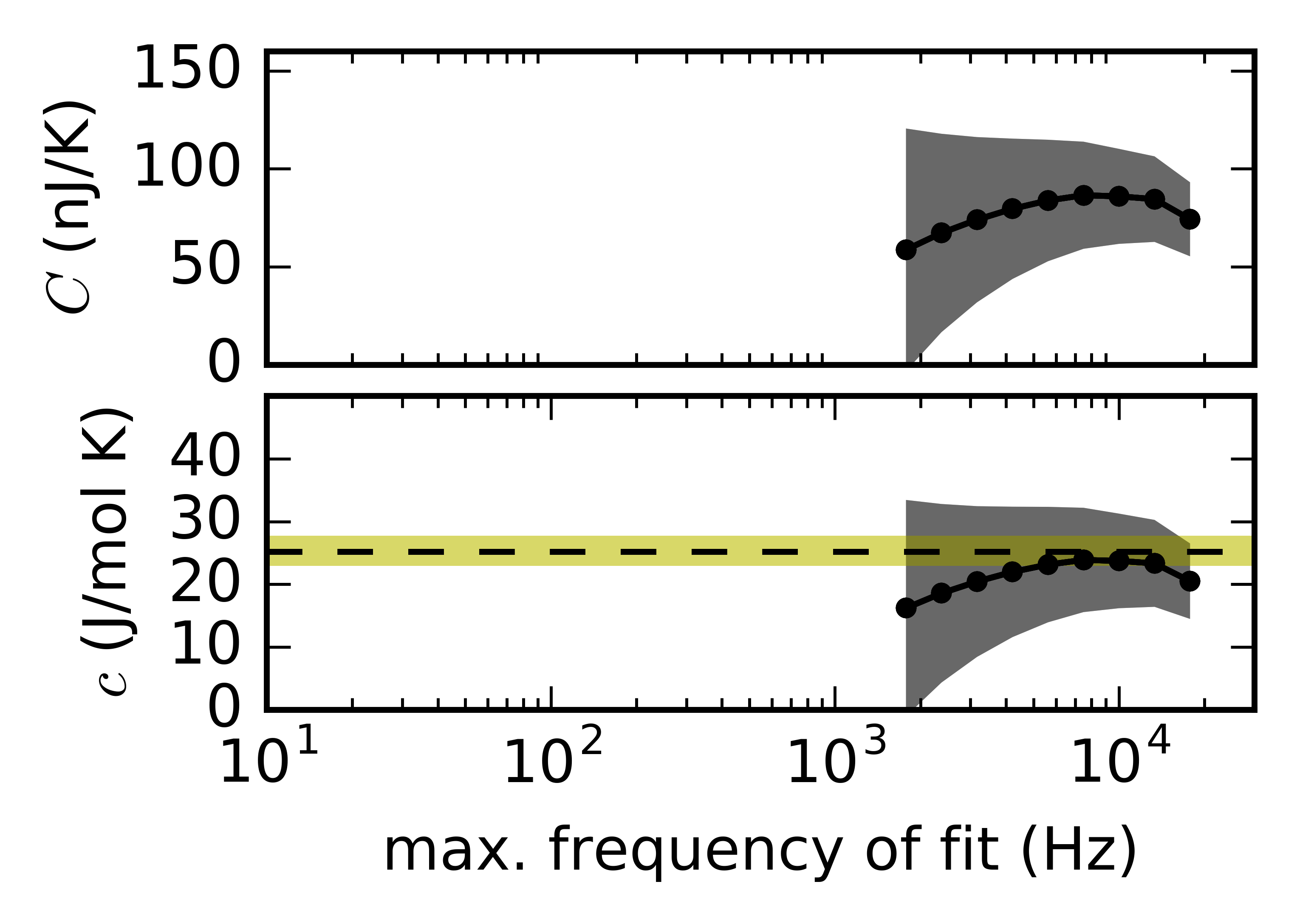}
\caption{Laboratory data and fitted heat capacities of an iron sample. The top three panels show measured third harmonic amplitudes, inferred temperature oscillations, and effective thermal conductance of the sample plus surrounding, plotted versus the frequency of driving voltage, $\omega/2\pi$. Blue, green, red and cyan indicate $\pm 9.6$, 7.8, 6.4, and 5.4 V driving voltages, respectively. Shaded areas reflect uncertainties. The solid black curve in the third panel shows the three-parameter fit using data from 10 Hz to 20 kHz, while dashed black lines show the contributions to $K_{\th}$ from each of the three terms in the model. Black circles in the bottom two panels show total sample heat capacity and specific heat of iron inferred from model fits that use data from 10 Hz to maximum frequencies of 2 kHz to 20 kHz, with uncertainties shown in grey. The dashed black line in the bottom panel shows the literature value of specific heat and yellow shades the values within $\pm 10\%$ of the literature value.}
\label{fig:Fe_data}
\end{center}
\end{figure}
	\begin{figure}
		\centering
		\includegraphics[width=3in]{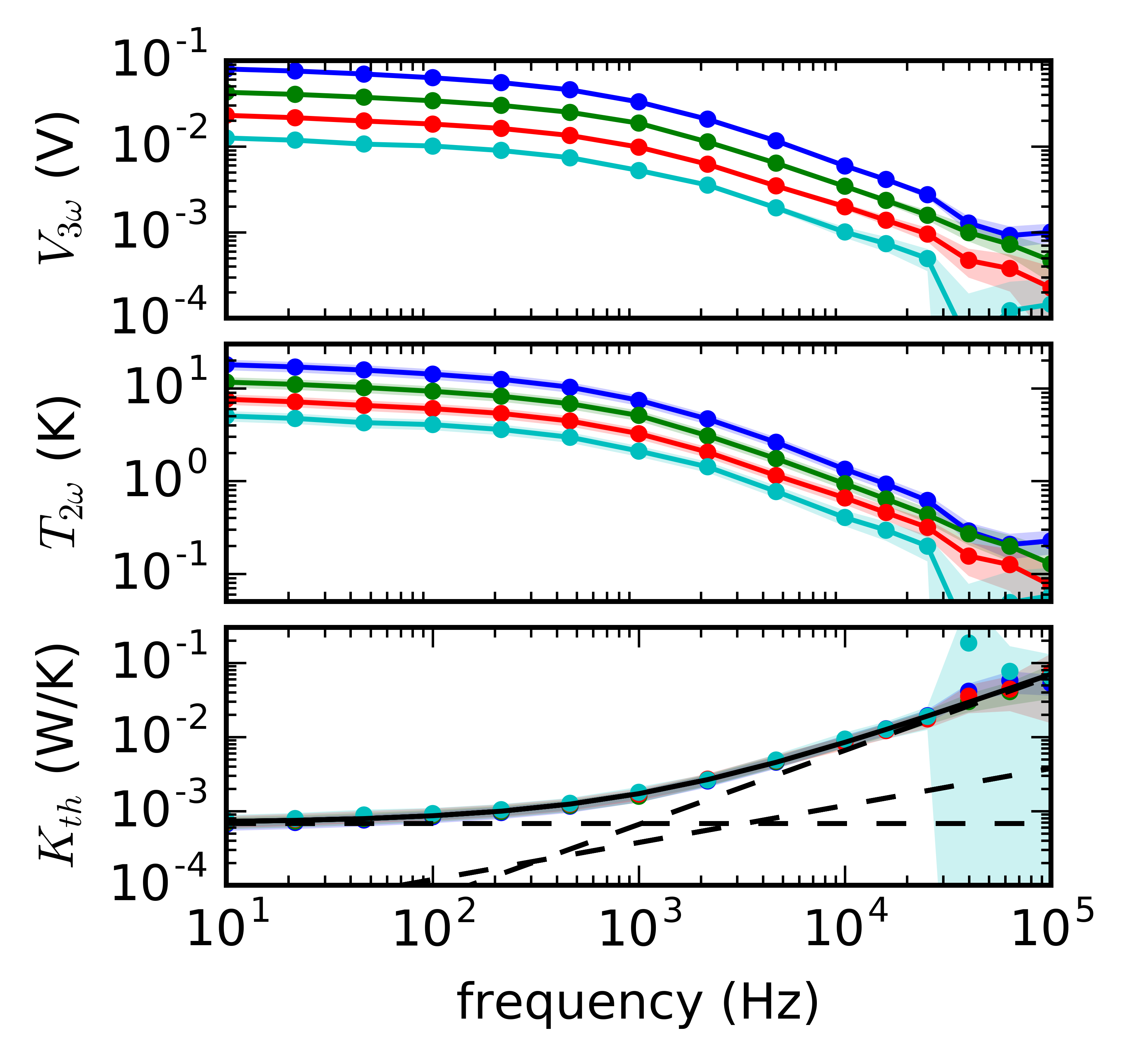} 
		\includegraphics[width=3in]{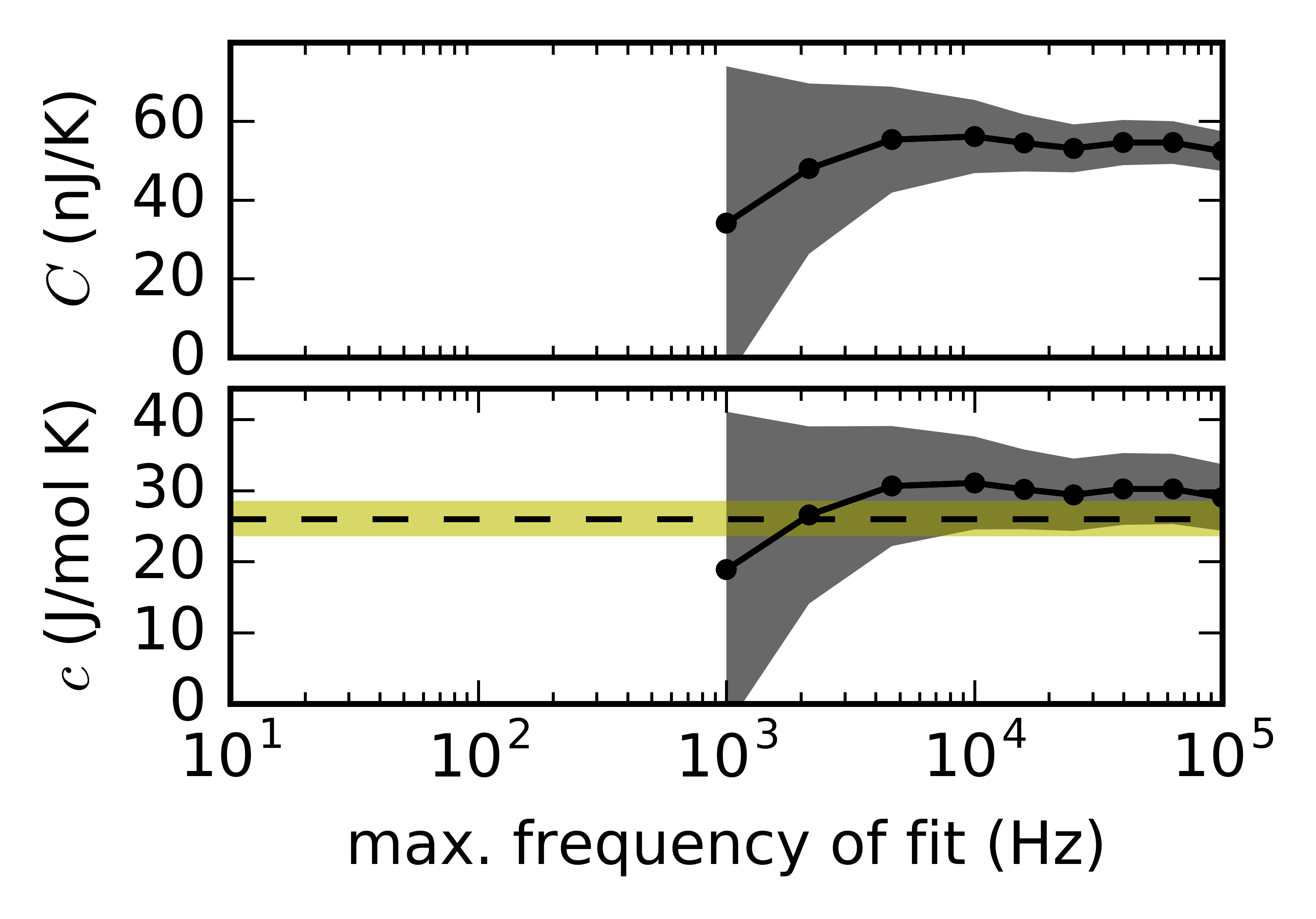}
		\caption{Same as Fig. \ref{fig:Fe_data}, but with the platinum foil replacing the iron sample, slightly different driving voltages ($\pm10$, 8.2, 6.8 and 5.6 V for blue, green, red and cyan circles) and 100 kHz bandwidth used for the fit shown in the third panel.}
		\label{fig:Pt_data}
	\end{figure}
\begin{figure}
	\includegraphics[width=3in]{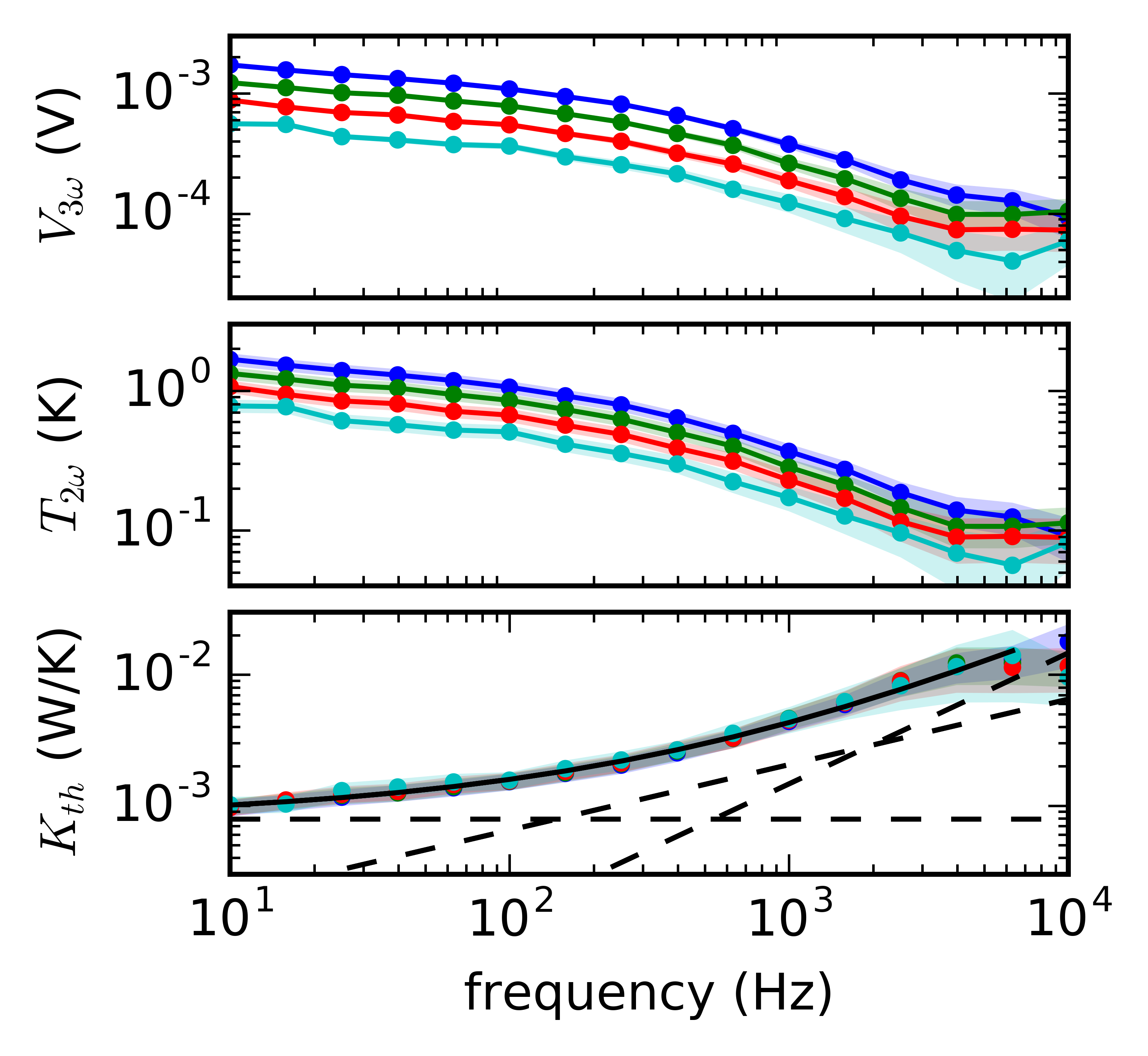} 
	\includegraphics[width=3in]{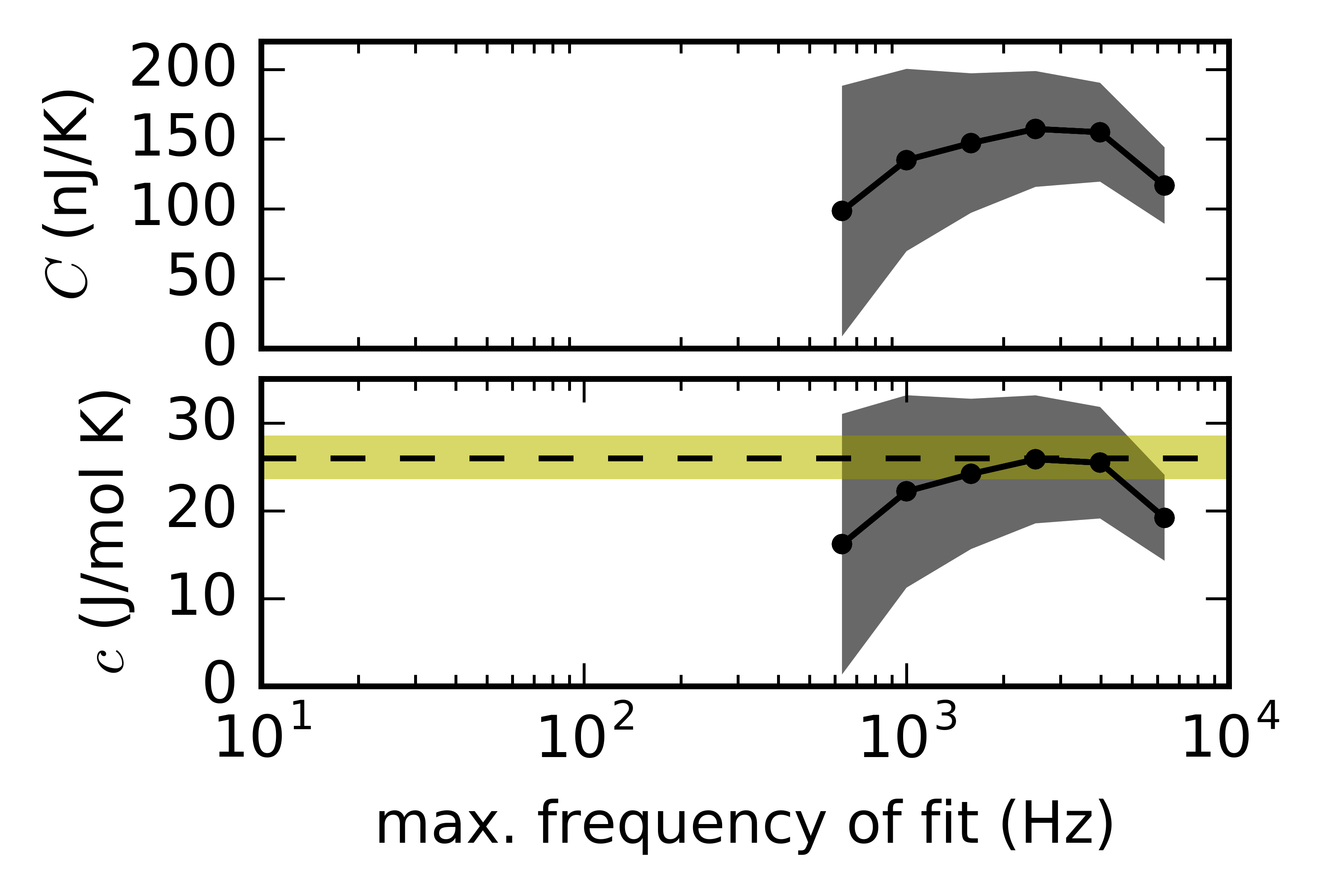}
		\caption{Same as Figs. \ref{fig:Fe_data}-\ref{fig:Pt_data}, but with a nickel foil instead of iron or platinum sample, slightly different driving voltages ($\pm10$, 9, 8 and 7 V for blue, green, red and cyan circles) and 7 kHz bandwidth used for fit shown in the third panel.}
		\label{fig:Ni_data}
	\end{figure}
		\begin{figure}
		\includegraphics[width=3in]{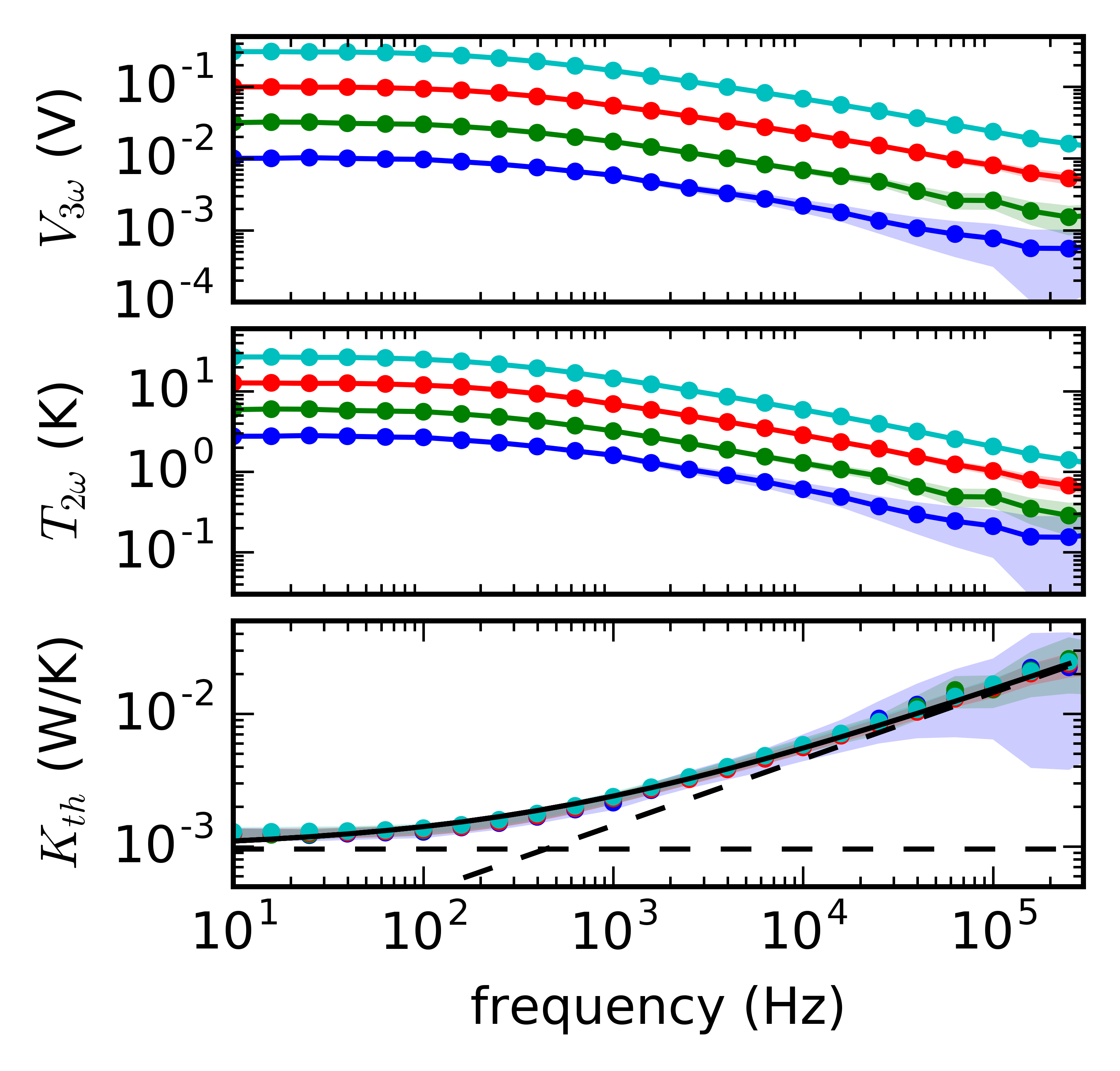} 
		\includegraphics[width=3in]{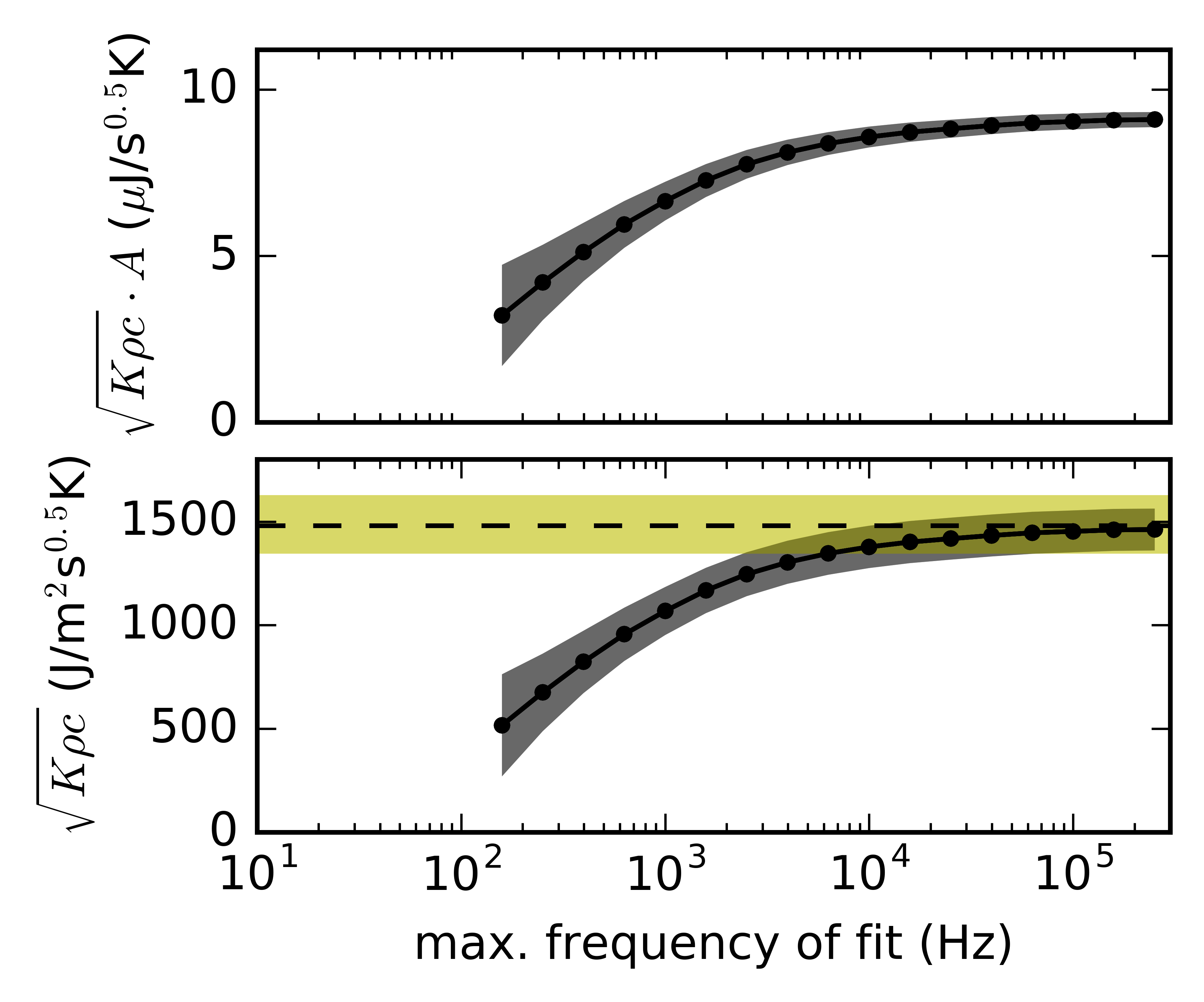}
		\caption{Laboratory data and fitted effusivities of a glass sample heated with a platinum thin-film. The top three panels show measured third harmonic amplitudes, inferred temperature oscillations, and effective thermal conductance of the heater plus surrounding, plotted versus the frequency of the driving voltage, $\omega/2\pi$. Blue, green, red and cyan indicate $\pm 6.3$, 4.3, 2.9, and 2.0 V driving voltages, respectively. Shaded areas reflect uncertainties. The solid black curve in the third panel shows the two-parameter fit using data from 10 Hz to 20 kHz, while dashed black lines show the contributions to $K_{\th}$ from the two terms in the model. Black circles in panels four and five show thermal effusivity times surface area (one of the fit parameters) and thermal effusivity of the glass insulation inferred from model fits that use data from 10 Hz to maximum frequencies of 200 Hz to 300 kHz, with uncertainties shown in grey. The dashed black line in panel five shows the literature value of thermal effusivity of silica glass and yellow shades the values within $\pm 10\%$ of the literature value.}
		\label{fig:glass_data}
\end{figure}

\section{Laboratory Results}
An example of laboratory data is shown in Fig. \ref{fig:data_ex}. As expected, the voltage measured across the bridge includes two components of roughly equal magnitude: a first harmonic and a third harmonic. The third harmonic amplitude, 4.1 mV, is used to infer a 1.1 K temperature oscillation following Eq. \ref{eqn:T2w_meas}, with all values of $R_\sam$ equal to the resistance of the sample section of the bridge circuit, $R_\sam^\textrm{2pt}$, except for the denominator, where $R_\sam = R_\sam^\textrm{4pt}$, the resistance of the sample alone, not including contact and lead resistance. \footnote{In this study, we measure $R_\sam^{2\textrm{pt}}$ and calculate $R_\sam^{4\textrm{pt}}$ based on measured dimensions of metal foils and the literature values of Fe, Pt and Ni resistivities, with one exception: for the thin-film Pt heater, we assume $R_\sam^{4\textrm{pt}}$ equals the measured value of $R_\sam^{2\textrm{pt}} = 26$ $\Omega$.}

We also vary frequency and driving voltage and plot the resulting third harmonic amplitudes and temperature oscillations in the first two panels of Fig. \ref{fig:Fe_data}. We then calculate the resistance of the system to changes in temperature, hereafter referred to as ``effective thermal conductance'' and denoted $K_\th$. It accounts for both the heat capacity of the metal sample and conductance to the surroundings, and is defined by:
\begin{equation}
K_\th = \frac{p_{2\omega}}{T_{2\omega}}
\end{equation} 
where $p_{2\omega} = \frac{1}{2}I_\omega^2R_{sam}^{4\textrm{pt}}$ and $I_\omega = V_d/(R_b+R_{sam}^{2\textrm{pt}})$.  In other words, it is the total heat capacity, $C_\total$, times the heating frequency, $2\omega$.

The top three panels of Fig. \ref{fig:Fe_data} show how measured third harmonics of voltage, inferred temperature oscillations, and effective thermal conductances vary with frequency and driving voltage. Since we show in section  \ref{section:errors} that uncertainties in $V_{3\omega}$ are approximately -80 dBc (i.e. 0.01\% of the voltage across the sample, multiplied by gain), we assume this value here, and propagate it, along with $15\%$ uncertainty in $R_\sam^\textrm{4pt}$, to uncertainties in $T_{2\omega}$ and $K_{\th}$. 

As heat input per cycle decreases due to increasing frequency or decreasing driving voltage, temperature oscillations decrease and so do third harmonic voltages. The fact that the effective thermal conductance, $K_{\th}$, is independent of driving voltage, means that a true third-harmonic is being measured (i.e. $V_{3\omega} \propto V_d^3$).

The shape of the $K_{\th}$ vs. frequency curve can be understood using a one-dimensional model of heat flow.\footnote{By fitting $K_{\th}$ vs. frequency, we use more information to infer $c_\sam$ than in the numerical models, enabling more accurate measurements, which is important given the relatively-limited bandwidth of laboratory measurements.}  We model $K_{\th}$, the heating rate required to raise the temperature of the metal by one degree, as the sum of three rates: (1) the rate required to raise the temperature of the metal alone, $c_\sam\rho_\sam\textrm{Vol} \cdot 2\omega$, (2) the rate required to heat the insulation via thermal diffusion, $\sqrt{k_\ins\rho_\ins c_\ins} \cdot \textrm{Area}\cdot \sqrt{2\omega}$, and (3) the rate required to maintain a linear temperature gradient to the diamond heat sink, $k_\ins\textrm{Area}/d_\ins$. These three contributions to the effective thermal conductance are shown as black dashed lines in Fig. \ref{fig:Fe_data}, with slopes of 1, $\frac{1}{2}$ and 0 in log-log space, and with their sum (i.e. the fit to data) shown as a solid black curve. Fitted parameters in this one-dimensional model provide qualitative measures of insulation thickness and effusivity, but a quantitatively meaningful value of sample heat capacity since the heat-capacity term is most sensitive to data at high-frequency, where the one-dimensional model mimics reality. The fit to $K_{\th}$ data from 10 Hz to 24 kHz at all driving voltages yields a total heat capacity of $74 \pm 19$ nJ/K, shown as a black dot with grey error envelope at 24 kHz in the fourth panel of Fig \ref{fig:Fe_data}. Fits are weighted using measurement uncertainty, and determined (along with covariance) via the ``curve fit'' function within the SciPy library in python. We also test how the maximum frequency of fitted data controls heat capacity estimates (fourth panel of Fig. \ref{fig:Fe_data}) and specific heat estimates (fifth panel). To estimate specific heat and its uncertainty, we divide fitted heat capacity by the number of moles in the $530 \times 8.5 \pm 1 \times 5.7 \pm 0.5$ $\mu$m piece of Fe, and add the $20\%$ uncertainty in sample volume to the total uncertainty (assuming uncertainties add in quadrature). Using data from 10 Hz to 20 kHz, for example, we estimate specific heat to be $20 \pm 9$ J/mol K, which is consistent with the literature value of 25 J/mol K.

The results of the same analysis on platinum and nickel foils of similar dimensions show similar results. A 350 ng-mass strip of platinum ($460 \times 15 \pm 1.5 \times 2.4 \pm 0.2$ $\mu$m) is found to have a specific heat of $29 \pm 5$ J/mol K (compared to the literature value of 26 J/mol K) by fitting data from 10 Hz to 100 kHz. Here, the higher bandwidth is enabled by the slightly smaller cross-sectional area of this foil compared to the iron foil. Measurements of a lower resistance strip of nickel almost matches the literature value within the uncertainty: by fitting data from 10 Hz to 10 kHz, a 350 ng-mass piece of nickel ($420 \times 16 \pm 1 \times 6$ $\mu$m) is found to have a specific heat of $20 \pm 5$, whereas the literature value is 26 J/mol K.\footnote{In the case of nickel, we did not measure sample thickness, but rather rely on the manufacturer estimate of 6 $\mu$m-thickness.}

We also test an alternative Joule-heating calorimetry experiment: measurement of thermal effusivity of the insulation surrounding a thin-film heater. The results are shown in Fig. \ref{fig:glass_data}, with the same data processing for $T_{2\omega}$ and $K_{\th}$ as in the case of metal foils. The model used to fit  $K_{\th}$, however, is slightly different. Rather than using a three parameter fit and interpreting the heat capacity term quantitatively, we assume the heat capacity term is negligible and that the thermal effusivity term is quantitatively meaningful. Despite not modeling the experiment here, we expect this procedure to give quantitative estimates of thermal effusivity since as frequency becomes large, the one-dimensional model mimics the reality of heat diffusing from a thin-film heater into the glass insulation. More precisely, our one-dimensional model is realistic when the heating timescale is short compared to timescale of thermal diffusion across the width of the thin-film heater, $\textrm{width}^2c_\ins\rho_\ins/k_\ins$, but long compared to the timescale of conduction out of the metal thin-film, $(d_\heater c_\heater\rho_\heater)^2/(c_\ins\rho_\ins k_\ins)$. For the experiment tested here, this requires the heating frequency, $2f$, to fall within the range $2.7 \textrm{ kHz} < 2f < 110$ MHz.

Indeed, the final panel of Fig. \ref{fig:glass_data} shows that by using data from 10 Hz to any maximum frequency between 10 kHz and 300 kHz, the fitted value of thermal effusivity matches the literature value within the $\pm 6\%$ uncertainty. The precision of the fit to the product of thermal effusivity and surface area is better, reaching 2.5\% when data up to 300 kHz is used.

\begin{figure}
	\begin{center}
	\includegraphics[width=3.5in]{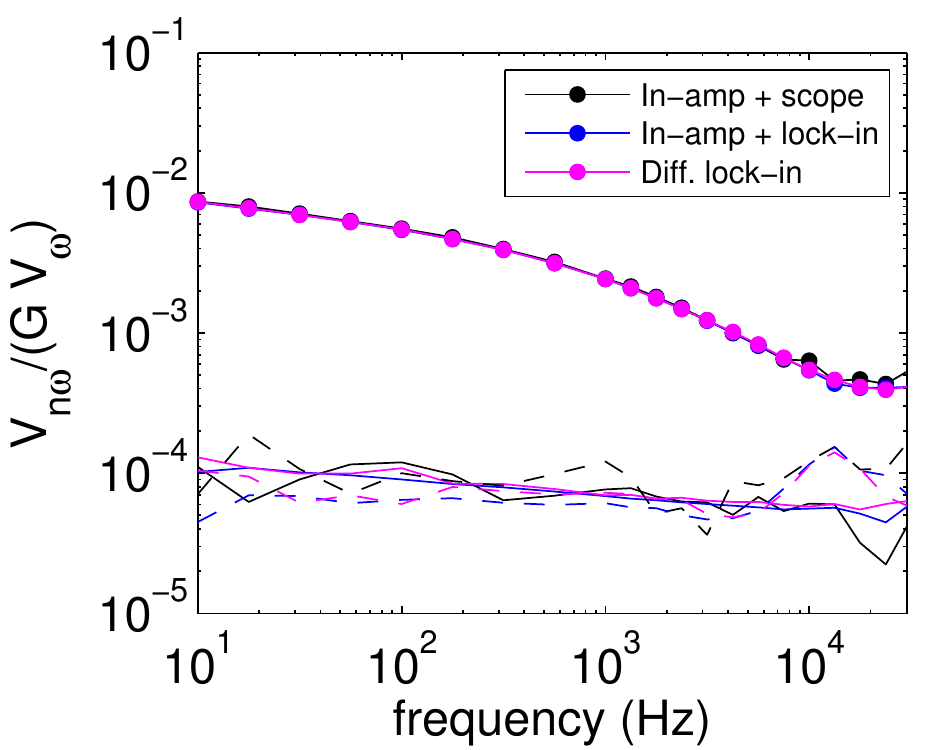} 
	\caption{Amplitudes of second and third harmonic voltages normalized by the product of gain and the first harmonic voltage oscillation across the sample or dummy. Circles represent third harmonics ($n=3$) of the strip of iron (Figs. 4, 6) under the greatest driving voltage ($\pm 9.6$ V), with colors indicating the electronics used to difference and digitize voltages (see legend). Dashed curves represent second harmonics of voltage ($n=2$) measured in the same experiments, a proxy for spurious third harmonics. Solid curves represent third harmonics ($n=3$) measured with the same electronics, but with a 1.5 $\Omega$ dummy sample (an off-the-shelf resistor), providing a second estimate of spurious third harmonics.}
	\label{fig:Fe_errors}
	\end{center}
\end{figure}
	\begin{figure}
		\includegraphics[width=3.5in]{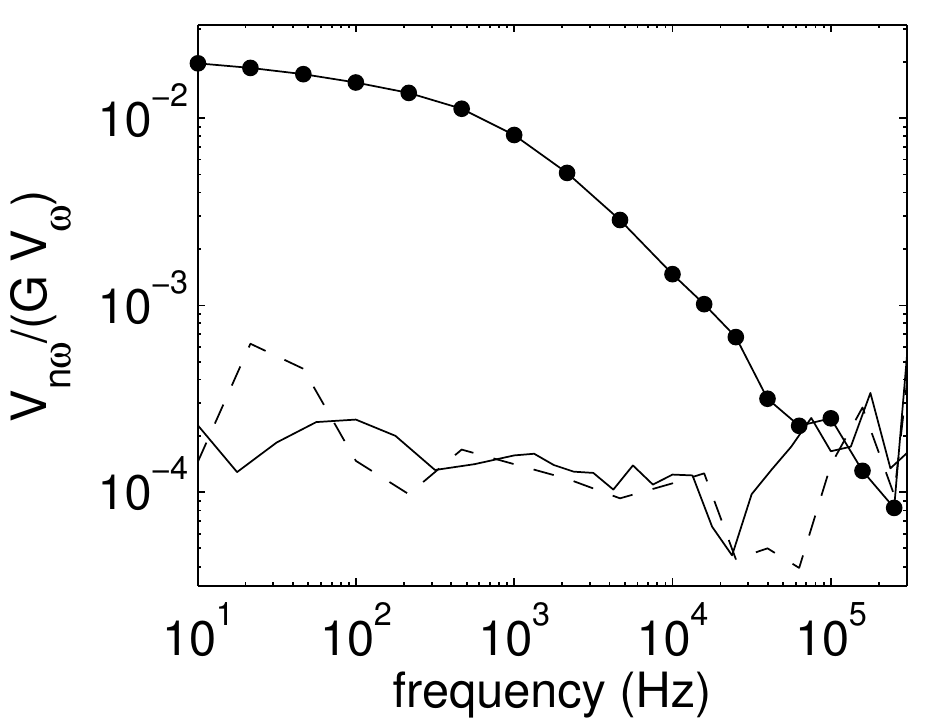} 
		\caption{Same as Fig. \ref{fig:Fe_errors}, but for the platinum foil, and using the ``in-amp + scope'' measurment scheme only.}
		\label{fig:Pt_errors}
	\end{figure}
\begin{figure}
		\includegraphics[width=3.5in]{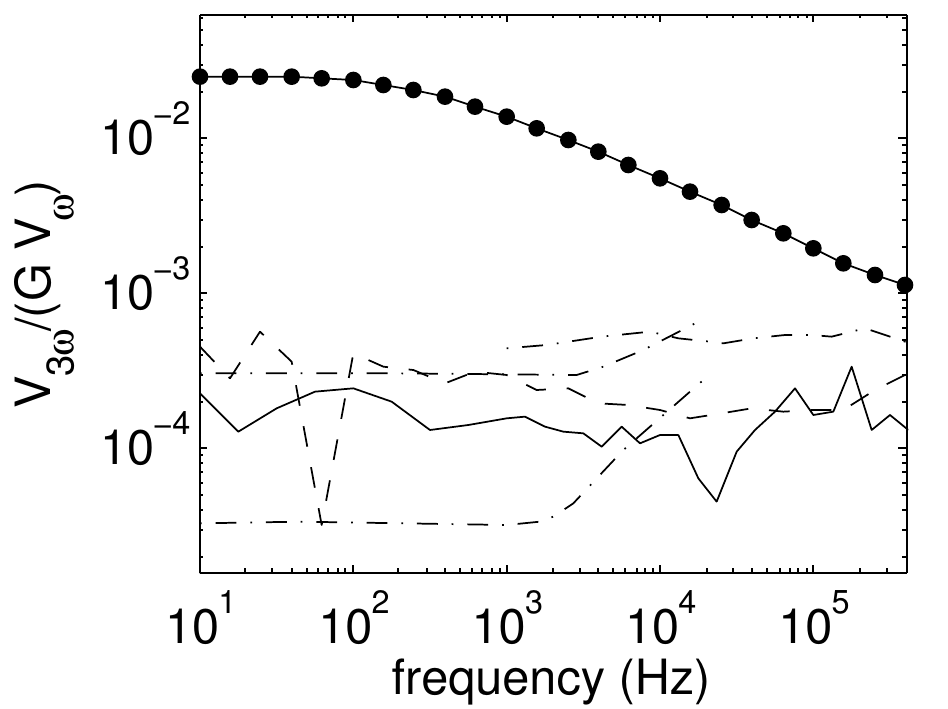} 
		\caption{Same as Figs. \ref{fig:Fe_errors},\ref{fig:Pt_errors}, but for the platinum thin film surrounded by glass, and with the largest bandwidth (up to 1 MHz).}
		\label{fig:glass_errors}
\end{figure}

\section{Error Analysis}
\label{section:errors}
To estimate the uncertainty in measured third harmonic amplitude, we measure background spurious harmonics in two ways: third harmonics of a dummy sample and \textit{second} harmonics of the real sample. The dummy sample is a 1.5 $\Omega$ off-the shelf resistor ($< 500$ ppm/$^\circ$C temperature coefficient, $1/4$ W power rating). 

Figs. \ref{fig:Fe_errors}-\ref{fig:glass_errors} show that both estimates of spurious harmonics imply a spurious free dynamic range of $80 \pm 10$ dBc. The plotted data is normalized by dividing by gain times the voltage measured across the sample, $V_\omega$. Third harmonics measured during real experiments (same as shown in Figs. \ref{fig:Fe_data}-\ref{fig:glass_data}) are also plotted for reference. They show that the signal starts $\sim 30$ dB above the spurious signal at low frequency, and approaches the spurious signal around 20 kHz for metal foils, whereas it approaches the spurious signal around 600 kHz in the case of the thin-film. 

The improved bandwidth for thin-films can be understood by the increase in thermally-induced third harmonic voltage for large values of $V_\omega$, the voltage oscillation across the heater. We expect the thermally-induced third harmonic to be proportional to $V_\omega^3$ whereas the spurious third harmonic are likely proportional to $V_\omega$ in our setup. $V_\omega$, in turn, is limited to the value $10 \textrm{V} \cdot R_\sam/(97 \Omega + R_\sam) \leq 150$ mV for the low-impedance metal foils (0.3 to 1.4 $\Omega$), but it is much larger for the high-impedance thin-film (26 $\Omega$). In practice, we limit the driving voltage to $\pm6.3$ V (compared to the $\pm10$ V possible from our waveform generator), but this still generates a 10-times larger voltage across the heater, compared to the metal foils.

Using the iron foil, we perform the tests of background spurious harmonics for three different voltage-measuring schemes mentioned above: instrumentation amplifier followed by oscilloscope, instrumentation amplifier followed by lock-in amplifier, and the lock-in amplifier alone (using its differential inputs). In all cases, we measure similar harmonics, showing that a lock-in amplifier does not improve the ratio of signal to spurious signal, as expected for harmonic distortions generated inside preamplifiers or waveform generators.


\section{Discussion}
Results of both the numerical and physical model of high-pressure experiments show a trade-off between bandwidth and measurement accuracy. In all cases, the trade-off is caused by spurious harmonics that overwhelm the third-harmonics induced from thermal oscillations in the sample. The coupled electrical-thermal numerical model shows that the bandwidth limit of our experiment is at most 3 MHz, at which point the instrumentation amplifier starts to generate overwhelming spurious harmonics. 

The physical model of diamond-cell experiments reveals a significantly more limited bandwidth in the case of low-impedance heaters. Third harmonic measurements are one order of magnitude above background (suggesting 10\% electrical error) at 1 to 10 kHz for $\sim 5 \times 13 \times 600$ $\mu$m metal strips. The larger impedance of the platinum thin-film allows for significantly more accurate electrical measurements at high frequency; the third harmonic measurement is one order of magnitude above background at 300 kHz.

Nonetheless, by fitting data on metal foils from 10 Hz to $\sim 10$ kHz to a three parameter model, we find heat capacities that agree with the literature value to within the $\pm 20$ to $30\%$ uncertainty of the fit.

The dominant source or sources of harmonic distortion (hence limited bandwidth) in our experiments may be the waveform generator, the voltage measuring devices, or both. In fact, the harmonic distortion expected in either differencing amplifier is in the range measured in our experiment: total harmonic distortion generated in the HF2LI under one-sided drive is, according to the manufacturer, approximately -70 dB, while the third harmonic distortion from the manufacturer of AD8429 (a newer version of AD8421) ranges from -90 dB to -60 dB, depending on frequency and gain.\footnote{See specifications for HF2LI and AD8429 at www.zhinst.com and www.analog.com } Still, we expect lower harmonics under the symmetric drive used here than the one-sided drive used to test the amplifiers, suggesting that the -80 dBc spurious signal may come from an alternative source.

The other likely source of spurious harmonics is the waveform generator. The manufacturer reports $< -54$ dB harmonic distortion, but this is reported as an upper limit and it is measured from one channel, so we expect smaller distortions when nearly-balanced waveforms from the two outputs are differenced, as in our setup. To discriminate harmonic distortions internal to the voltage generating unit from those internal to the voltage measuring unit, lower-distortion test equipment would be needed. In fact, analog audio analyzers may provide lower harmonic distortion that would enable such tests. For example, the Keysight U8903A is reported to have total harmonic distortion $\leq -101$ dB at 20 Hz to 20 kHz \footnote{See specifications for U8903A at www.keysight.com}. Use of such a device in a real experiment could also improve measurement accuracy.

But harmonic distortion is not the only source of uncertainty in our measurement. Uncertainty in sample resistance is a key error source in the measurements of metal foils because we only measured the two-point resistance of the sample plus leads and contacts, leaving uncertainty in the theoretically calculated sample resistance. In fact, this $\sim \pm 10\%$ uncertainty propagates to $\pm 20\%$ uncertainty in $T_{2\omega}$ and $K_{\th}$. In the case of a thin-film heater with low contact resistance, the two-point resistance measurement, $R_\sam^{\textrm{2pt}}$, is assumed to be an accurate measure of the true sample resistance, $R_\sam^{\textrm{4pt}}$.

Finally, the uncertainty estimate for specific heat or thermal effusivity includes a significant contribution from uncertainty in heater size. Here we use laser cutting, cutting by hand with a razor blade, and photolithography of a poorly adhered platinum film. These techniques result in somewhat rough edges, resulting in uncertainties of $\pm 1$ $\mu$m or $\sim 10 \%$ in width, which propagates to $10 \%$ in specific heat or effusivity. 

In principle, the uncertainties in heater resistance and dimensions can be greatly reduced by use of advanced fabrication techniques such as focused-ion-beam milling, mechanical cutting with a micromanipulator, or photolithography. Altogether, the improvements outlined here could enable measurements with 1\% to 10\% accuracy, as modeled numerically.

But using the current setup with the current error sources, quantitative high-pressure measurements of heat capacity or thermal effusivity are already possible if the appropriate heaters, samples and pressure devices are employed. The design must simply allow for the uncertainties measured here: -80 dBc harmonic purity at $\pm 10$ V driving voltage, and $\pm 1$ $\mu$m sample dimensions.

In a diamond-cell capable of reaching 30 GPa (using anvils with $\sim 500$ $\mu$m-diameter culets), this would enable heat capacity measurements of metals with slightly lower accuracy than those presented here, because the sample would have to be slightly shorter than the ones studied here. Heat capacity measurements of semimetals or semiconductors, on the other hand, would reach significantly higher accuracy if they were shaped in a way that their impedances were 10 to 100 $\Omega$. 

The measurements of thermal effusivity presented here are from samples loaded into the 300 $\mu$m-diameter sample chamber of a diamond cell capable of reaching $\sim 20$ GPa. Hence, it is possible that diamond-cell experiments to  $> 10$ GPa could reach the $\pm 6\%$ accuracy documented here for thermal effusivity. One significant challenge in performing such high-pressure measurements is to avoid breaking the thin-film heater upon application of pressure. 

In a larger volume presure cell, the uncertainty in heat capacity could be significantly reduced, provided the $\sim 5 \times 13$ $\mu$m-cross sectional area could be maintained while increasing the length of metal sample. Specifically, we expect uncertainty in $V_{3\omega}$ to be inversely proportional to length if all other parameters are fixed, in this case of a low-impedance heater. Alternatively, a thin-film heater (or other heater with $\sim 10$ to 100 $\Omega$ impedance) would allow quantitative measures of effusivity, as in the case of diamond-cell experiments. 

Analogous measurements of as-grown materials on high thermal conductivity substrates should follow the same guidelines: using the electrical test equipment employed here, $\sim 1$ $\Omega$ Joule-heaters will allow measurements up to $\sim 10$ kHz, while 10 to 100 $\Omega$ heaters will allow measurements up to 300 kHz frequency. Possible applications include direct measurement of heat capacity of $\sim 5 \times 13 \times 600$ $\mu$m-thick semiconductors grown on sapphire substrates (e.g. GaN \cite{Paskova1999}), or as in the case of the high-aspect ratio metal sample proposed above, a metal sample of similar cross-section and ten-times greater length (e.g. MgB$_2$ \cite{Moeckly2006}). 

Finally, we note that the accuracy needs of measurements depend greatly upon the scientific or engineering question to be addressed. For example, the 10\% accuracy threshold assumed here is not likely to be useful for detailed thermodynamic analysis, but is sufficient for detection of a wide variety of second-order phase transitions. Moreover, progress in sample preparation (to reduce uncertainties in volume) and electrical test equipment (to increase bandwidth and/or decrease harmonic distortion) may lead $\ll 10\%$ accuracy in future heat capacity measurements.

\section{Conclusions}
Physical models of high-pressure Joule-heating calorimetry experiments show that heat capacity can be measured directly with $20$ to $30\%$ accuracy, while thermal effusivity can be measured with $6\%$ accuracy. Harmonic distortions due to electrical test equipment cause these uncertainties to be larger than those estimated from numerical models.  Nonetheless, the current experimental setup may enable a wide variety of experiments in high-pressure science. 


%
%

\appendix
		
\section{}
We derive an approximate analytic solution to the steady state heat equation in order to calculate the resistance needed to balance the electrical bridge in our coupled electrical-thermal numerical model.

The voltage source, $V_d \sin(\omega t)$, drives current through a series of resistors with total initial resistance $R_{ti}$, including the sample, which heats up. The sample's temperature increases to a higher steady state value, $T_{DC}$, and oscillates, $T_{AC}(t)$, causing an increased total resistance, $R_{ti}+ R_0\alpha (T_{DC}+T_{AC})$, where $R_0$ is the ambient temperature sample resistance and $\alpha = d\textrm{log}R/dT$ is the sample's temperature coefficient of resistance. A thermal link to a constant-temperature reservoir (e.g. the diamonds) with thermal conductance $K_{\th}$ (units: W/K) cools the sample. For simplicity in this steady state calculation, we assume the insulation's heat capacity is zero, meaning the change in temperature of the sample is due to two terms only: Joule heating, $I^2R$, and heat conducted away, $K_{\th}T$. The current, $I$, is a ratio of driving voltage to time-dependent total resistance: 
\[ I = \frac{V_d \sin(\omega t)}{R_{ti} + \alpha(T_{DC}+T_{AC})R_0}
\]

resulting in the following heat equation:

\begin{eqnarray}
C_{\sam}\frac{dT_{AC}}{dt} &=& I^2 R - K_{\th}T \nonumber \\
&=& \frac{V_d^2\sin^2(\omega t)R_0(1+\alpha(T_{DC}+T_{AC}))}{(R_{ti} + \alpha(T_{DC}+T_{AC})R_0)^2} \nonumber \\
&-& K_{\th}(T_{DC}+T_{AC}) 
\label{eqn:heat}
\end{eqnarray}
.

To solve for the steady state temperature, we eliminate the sinusoidal terms, assume temperature oscillations are small ($\frac{T_{AC}}{T_{DC}} \ll 1$), and use the identity $\sin^2(\omega t) = \frac{1}{2}(1 - \cos(2\omega t))$ to arrive at the expression:

\[
0 = \frac{\frac{1}{2}V_d^2R_0(1+\alpha T_{DC})}{(R_{ti} + \alpha T_{DC} R_0)^2} - K_{\th}T_{DC}
\label{eqn:heat_DC}
\]

Dividing by $-K_{\th}$ and rearranging,
\[ 0 = T_{DC} - \frac{V_d^2R_0(1+\alpha T_{DC})}{2K_{\th}(R_{ti} + \alpha T_{DC} R_0)^2}
\]

Multiplying by $(R_{ti} + \alpha T_{DC} R_0)^2$, expanding and grouping into powers of $T_{DC}$ results in the following cubic equation:

\begin{eqnarray*}
0 = (\alpha R_0)^2 T_{DC}^3 &+& 2\alpha R_0R_{ti} T_{DC}^2 + \left(R_{ti}^2 - \frac{V_d^2\alpha R_0}{2K_{\th}}\right)T_{DC} \\
& -& \frac{V_d^2R_0}{2K_{\th}}
\end{eqnarray*}

To find the steady state temperature rise, we use the cubic formula, and assume the maximal root is the correct solution. We confirm the correct choice of cuibc root by running the numerical simulation itself.

\section{}

Here we derive the capacitances, resistances, and current sources needed to implement our one-dimensional heat flow model in LTSpice. Part I of this two-part publication explains the reduction of the heat equation to one dimension in our planar model:
\[ \frac{\partial T}{\partial t} = \frac{1}{\rho c} \left( k \frac{\partial ^2 T}{\partial z^2} + p \right)
\] 
where $k$, $c$, and $\rho$ are material properties listed in Table 6.1, $z$ is the axial direction in a diamond cell, and $p$ is power density.

\begin{widetext}

Discretizing, rearranging, and multiplying by $\frac{A}{A}$ where $A$ is the surface area of the metal sample, we find,

\begin{equation}
\frac{T_i^{n+1}-T_i^{n}}{\Delta t} = \frac{1}{\rho c A\Delta z} \left( \frac{k_{i+0.5}A(T_{i+1}-T_i)-k_{i-0.5}A(T_i-T_{i-1})}{\Delta z}  + p_iA\Delta z \right)
\label{eqn:thermal_discrete}
\end{equation}
where the subscript marks the position in space (with fractions indicating the value of the link between two elements), and the superscript marks the position in time.

In the LTSpice implementation of the electrical schematic shown in Fig. \ref{fig:electrical-thermal}, the rate of change of voltage across capacitor $i$ at time $n$ is current-in minus current-out, and current is given by the negative gradient of voltage divided by resistance: 
\begin{eqnarray}
\frac{V_i^{n+1} - V_i^{n}}{\Delta t} &=& \frac{I_{i-0.5} - I_{i+0.5}+I_i^\source}{C_i} \nonumber \\
&=& \frac{1}{C_i}\left( \frac{V_{i-1}-V_i}{R_{i-0.5}} -\frac{V_{i+1}-V_{i-1}}{R_{i+0.5}} + I_i^\source \right) 
\label{eqn:electrical_discrete}
\end{eqnarray}
where the fractional subscript $i+0.5$ indicates the resistor that separates capacitors $i$ and $i+1$.

\end{widetext}

Comparison of Eqs. (\ref{eqn:thermal_discrete}) and (\ref{eqn:electrical_discrete}) shows that the analogies listed in Table \ref{table:electrical-thermal} are valid: 
\begin{eqnarray}
V_i &\longleftrightarrow& T_i \nonumber \\
C_i &\longleftrightarrow& \rho c  A\Delta z \nonumber \\
R_i &\longleftrightarrow& \frac{\Delta z}{k_i A} \nonumber \\
I_i^\source  &\longleftrightarrow&  p_i A\Delta z  \nonumber
\end{eqnarray}

To improve accuracy, we average models using two alternative values of resistance at the interface between insulation and sample: $R_\textrm{inter} = \frac{2\Delta z}{k_{\ins} + k_{\sam}}$ and $R_\textrm{inter} = R_\ins$. The difference between temperature oscillations inferred from the two models is 20\% over the range plotted in Fig. \ref{fig:CbyC_LTSpice}, but the average value is within 2\% of the numerically-accurate thermal model of Part I of this two-part publication.

\section{}
Here we calculate the amplitude of temperature oscillations, $T_{2\omega}$, and the heat capacity, $C_\total$, to be inferred from a measurement of third harmonic voltage, $V_{3\omega}$. First we calculate the more intuitive relationship, $V_{3\omega}$ as a function of $T_{2\omega}$, and then invert for the desired formula. Let $R_{\sam}$ be the time-averaged resistance of sample and let $R_b$ be the other resistors in the sample arm of the electrical bridge. Note that the sample's resistance should be measured using a DC or first harmonic measurement. In principle, $\alpha$ should also be measured by varying temperature slightly during a DC or other low frequency resistance measurement, though in practice we assume literature values for the foils of iron, platinum and nickel.

\begin{widetext}

The temperature oscillation causes a resistance oscillation in the sample:
\[
R_{\sam}(t) \approx R_{\sam} (1+\alpha T_{2\omega} \sin(2\omega t+\phi))
\]
where $\phi$ is a phase shift that accounts for the possibility that the cosine component is not negligible.

The total voltage oscillation across the sample is therefore,
\begin{eqnarray}
V_\sam &\approx& \frac{V_d \sin(\omega t)}{R_b+R_\sam+R_\sam \alpha T_{2\omega}\sin(2\omega t+\phi)}R_\sam(1+ \alpha T_{2\omega} \sin(2\omega t+\phi)) \nonumber \\
&=& \frac{V_d R_\sam}{R_b + R_\sam} \left( \frac{\sin(\omega t ) (1 + \alpha T_{2\omega} \sin(2\omega t+\phi)) }{1+\frac{R_\sam}{R_b+R_\sam}\alpha T_{2\omega} \sin(2\omega t+\phi)} \right) \nonumber \\
&\approx& \frac{V_d R_\sam }{R_b + R_\sam} \sin(\omega t)(1 + \alpha T_{2\omega} \sin(2\omega t+\phi)) \left( 1 - \frac{R_{\sam}}{R_b + R_\sam} \alpha T_{2\omega} \sin(2\omega t+\phi)\right) \nonumber \\
&=& \frac{V_d R_\sam }{R_b + R_\sam} \sin(\omega t) \bigg[ 1 + ( 1 - \frac{R_\sam}{R_b + R_\sam} ) \alpha T_{2\omega} \sin(2\omega t+\phi) + O((\alpha T_{2\omega})^2) \bigg] \nonumber
\end{eqnarray}
where the additional approximation (line 3) is that $\frac{1}{1+x} \approx 1-x$ for small $x$. The final term in line 4 is assumed to be small, and the first term is nulled out by a well-balanced electrical bridge, leaving a residual voltage across the bridge of,
\begin{eqnarray}
V_\resid &\approx& \frac{V_d R_\sam }{R_b + R_\sam} \left( 1 - \frac{R_\sam}{R_b + R_\sam}     \right) \alpha T_{2\omega}  \sin(\omega t) \sin(2\omega t+\phi) \nonumber \\
& = & \frac{V_d R_\sam R_b \alpha T_{2\omega}} {\left( R_b + R_\sam  \right)^2 } \frac{1}{2} \Big( \cos(\omega t+\phi/2)-\cos(3\omega t + 3\phi/2) \Big) \nonumber
\end{eqnarray}

The third harmonic amplitude of the expected residual voltage after being amplified by gain $G$ is therefore, 
\[
V_{3\omega} = \frac{GV_d R_\sam R_b \alpha T_{2\omega}}{2 \left( R_b + R_\sam  \right)^2 } 
\]

Inverting to solve for the temperature we derive Eq. (\ref{eqn:T2w_meas}), 
\begin{equation}
T_{2\omega} = \frac{2 \left( R_b + R_\sam  \right)^2 V_{3\omega}}{\alpha V_d R_\sam R_b G} 
\end{equation}

The total heat capacity of sample plus addenda is therefore,

\begin{eqnarray}
C_{\total} &=& p_{2\omega}/2\omega T_{2\omega} \nonumber \\
&=& \frac{1}{2} ( \frac{V_d}{R_\sam+R_b} )^2 R_\sam \frac{1}{2\omega T_{2\omega}} \nonumber \\
&=& \frac{ \alpha V_d^3 R_\sam^2 R_bG}{8 \omega (R_\sam + R_b)^4 V_{3\omega} } \\ 
\end{eqnarray}

\end{widetext}

\begin{acknowledgements}
We thank Ali Niknejad, Paul Goldey, and Norman Birge for advice in designing the electronics. Support for Z.M.G. was provided by CDAC and the Carnegie Institution for Science.
\end{acknowledgements}

\nocite{*}
\bibliography{AC_Cal2_v2}

\end{document}